\documentclass[aps,prd,groupedaddress]{revtex4}
\usepackage{graphicx}
\usepackage{epsfig}
\usepackage{amsfonts}
\usepackage{amssymb}
\usepackage{epsf}
\usepackage{xcolor}
\newcommand{\insertplot}[5]{\begin{figure}
 \hfill\hbox to 0.05in{\vbox to #5in{\vfill
 \inputplot{#1}{#4}{#5}}\hfill}
 \hfill\vspace{-.1in}
 \caption{#2}\label{#3}
 \end{figure}}
\newcommand{\inputplot}[3]{
 \special{ps: plotfile #1}
}

\newcounter{fig}

\begin{document}

\title{
Particle-like ultracompact objects
in Einstein-scalar-Gauss-Bonnet theories}

\author{Burkhard Kleihaus}
\email[]{b.kleihaus@uni-oldenburg.de}
\author{Jutta Kunz}
\email[]{jutta.kunz@uni-oldenburg.de}
\affiliation{Institut f\"ur Physik, Universit\"at Oldenburg, D-26111 Oldenburg, Germany}
\author{Panagiota Kanti}
\email[]{pkanti@cc.uoi.gr}
\affiliation{Division of Theoretical Physics, Department of Physics,
             University of Ioannina, GR-45110, Greece}

\date{\today}
\begin{abstract}
We present a new type of \textit{ultracompact objects},
featuring lightrings and echoes in the gravitational-wave spectrum.
These particle-like solutions arise
in Einstein-scalar-Gauss-Bonnet theories in four spacetime dimensions,
representing globally regular spacetime manifolds.
The scalar field diverges at the center, but the effective
stress-energy tensor is free from pathologies.
We determine their domain of existence and compare with wormhole solutions,
black holes and the Fisher solution. 
\end{abstract}
\maketitle
\section{Introduction}
%
In recent years, after the detection of gravitational waves 
by the LIGO-Virgo collaboration \cite{LIGO,VIRGO}
and the first image of a supermassive black hole 
by the EHT collaboration \cite{EHT}
the interest in compact objects as well as in their 
potential observable signatures has been greatly re-intensified.
At the same time this has led to a surge of interest
in horizonless compact objects and their properties.
Here we present gravitational particle-like solutions
that may indeed be classified as
\textit{ultracompact objects} (UCOs) \cite{Cardoso:2017cqb},
since they may possess lightrings.
At the same
time, being horizonless, they will feature a distinctive sequence of echoes
\cite{Cardoso:2017cqb,Cardoso:2016rao,Cardoso:2016oxy} in the waveform
of gravitational waves propagating in their vicinity.

Observing compact objects in the universe provides also a useful test-bed
for generalized gravitational theories as strong-gravity effects may reveal modifications
from General Relativity. Here, we focus on a theory that is motivated from a quantum
gravity perspective and includes higher-curvature contributions in the form of the quadratic
Gauss-Bonnet (GB) term \cite{Zwiebach:1985uq,Gross:1986mw,Metsaev:1987zx}.
In four space-time dimensions, the coupling to a scalar field 
then leads to a class of Einstein-scalar-Gauss-Bonnet (EsGB) theories,
characterized by the choice of the coupling function.
EsGB black holes and their properties have been widely studied already,
revealing new phenomena, like a minimum mass \cite{Kanti:1995vq}, 
the violation of the Kerr bound \cite{Kleihaus:2011tg}
and the presence of scalarization for a variety of forms of the coupling function
\cite{Sotiriou-Zhou,Antoniou:2017acq,Doneva:2017bvd,Silva:2017uqg,
Blazquez-Salcedo:2018jnn,Cunha:2019dwb,Collodel:2019kkx}.

The presence of the GB term in the theory leads to an 
{\it effective} stress energy tensor that may violate the energy conditions.
Interestingly, such EsGB theories then give rise to wormhole solutions,
that -- in stark contrast to General Relativity (GR) --
do not need any form of exotic matter
\cite{Kanti:2011jz,Kanti:2011yv,Antoniou:2019awm}.
Therefore, the presence of this plethora of new phenomena arising in 
EsGB theories makes one wonder whether gravitational particle-like solutions
featuring a completely regular space-time should not be another astounding
feature of these theories.

Here we show that this is indeed the case, by constructing numerous families
of particle-like solutions, that possess a globally regular 
asymptotically-flat metric and an everywhere regular
effective stress energy tensor.
We explore the solution space of these particle-like solutions, and address
their relation to wormholes, black holes, and the Fisher solution
\cite{Fisher:1948yn,Janis:1968zz,Wyman:1981bd,Agnese:1985xj,Roberts:1989sk}
 thus providing a complete characterization
of all static, asymptotically-flat, spherically-symmetric solutions in EsGB theory.
Finally, we scrutinize the novel particle-like solutions 
with respect to their potential astrophysical implications.
In particular, we show that many of them are highly compact, 
and may be classified as ultracompact objects, since 
they possess lightrings
and feature echoes in the gravitational-wave spectrum.

\section{EsGB Theory}

%
A general EsGB theory may be formulated in terms of the following 
effective action
\begin{eqnarray}  
S=\frac{1}{16 \pi}\int d^4x \sqrt{-g} \left[R - \frac{1}{2}
 \partial_\mu \phi \,\partial^\mu \phi
 + F(\phi) R^2_{\rm GB}   \right],
\label{act}
\end{eqnarray}
with a scalar field $\phi$, a coupling function $F(\phi)$,
and the GB invariant
$R^2_{\rm GB} = R_{\mu\nu\rho\sigma} R^{\mu\nu\rho\sigma}
- 4 R_{\mu\nu} R^{\mu\nu} + R^2$.

Variation of the action with respect to the metric and
scalar field
leads to the coupled set of field equations
\begin{equation}
G_{\mu\nu} =  T_{\mu\nu} \ , \ \ \ 
\nabla^2 \phi + \dot{F}(\phi)R^2_{\rm GB}  =  0 \ , 
\label{einfldeq}
\end{equation}
with effective stress energy tensor
\begin{equation}
T_{\mu\nu} =
-\frac{1}{4}g_{\mu\nu}\partial_\rho \phi \partial^\rho \phi 
+\frac{1}{2} \partial_\mu \phi \partial_\nu \phi
-\frac{1}{2}\left(g_{\rho\mu}g_{\lambda\nu}+g_{\lambda\mu}g_{\rho\nu}\right)
\eta^{\kappa\lambda\alpha\beta}\tilde{R}^{\rho\gamma}_{\alpha\beta}\nabla_\gamma \partial_\kappa F(\phi) \ .
\label{tmunu}
\end{equation}
Here, we have denoted the derivative with respect to the scalar field
by a dot, and we have defined 
$\tilde{R}^{\rho\gamma}_{\alpha\beta}=\eta^{\rho\gamma\sigma\tau}
R_{\sigma\tau\alpha\beta}$ and $\eta^{\rho\gamma\sigma\tau}= 
\epsilon^{\rho\gamma\sigma\tau}/\sqrt{-g}$.

Focussing on static spherically-symmetric solutions, we employ
the line-element 
\begin{equation}
ds^2 = -e^{f_0} dt^2 +e^{f_1}\left[dr^2 
+r^2\left( d\theta^2+\sin^2\theta d\varphi^2\right) \right],
\label{met}
\end{equation}
where $f_0$ and $f_1$ are functions of $r$ only; likewise
the scalar field depends only on $r$.
Substitution of the ansatz for the metric and scalar field in
the field equations yields a system of four coupled, nonlinear, 
ordinary differential equations. Three of them are of second order
and can be reduced to a system of two independent second-order equations for $f_1$ 
and $\phi$.
%
%
The fourth equation is of first order and serves as a constraint that determines
the form of $f_0$ in terms of $f_1$ and $\phi$. 
%
%
%

\section{Expansions}

%
%
We have performed an asymptotic expansion for $r \to \infty$ 
up to ${\cal O}\left(r^{-5}\right)$, showing asymptotic flatness,
and yielding the mass $M$ and the scalar charge $D$
of the solutions 
\begin{equation}
f_0 = - 2M/r + {\cal O}\left(r^{-3}\right) \ , \ \ \
f_1 = 2M/r- (D^2+4M^2)/8r^2 + {\cal O}\left(r^{-3}\right) \ , \ \ \
\phi = \phi_\infty -D/r +  {\cal O}\left(r^{-3}\right) \ .
\label{asym_exp}
\end{equation}

The expansion at the center is more involved, however, it
demonstrates the regularity of space-time at $r=0$ in all cases
we studied.
Here, we restrict to polynomial coupling functions 
$F(\phi) = \alpha \phi^n$, $n\geq 2$,
and dilatonic coupling functions $F(\phi) = \alpha e^{-\gamma \phi}$.
For a polynomial coupling function, as $r\to 0$, 
we assume power-series expansions for the metric functions 
while the scalar field behaves like $\phi =\phi_c -c_0/r +P_\phi(r)$,
where $P_\phi(r)$ is a polynomial. 
Substituting the expansions in the Einstein and scalar field equations,
the expansion coefficients can be determined successively.
It turns out that the lowest (non-zero) order is exactly $n$ 
in the expansion of the metric functions and $n-1$ in the expansion 
of $P_\phi(r)$. For instance, for $n=2$, we explicitly find  
\begin{equation}
f_0 = 
     f_{0c} 
     +\frac{e^{f_{1c}}}{64\alpha}\,r^2
     \left(1+\frac{2}{3}\frac{\phi_c}{c_0}r\right)
     +{\cal O}(r^4) \ , \ \ \ 
f_1  = 
     f_{1c} 
     +\left(\frac{e^{f_{1c}}}{64\alpha}
     +\frac{\nu_3}{6}\,r\right)\,r^2
     +{\cal O}(r^4) \ ,
\label{expf01c}   
\end{equation}

\vskip -0.4cm
\begin{equation}
\phi =  
      -\frac{c_0}{r}
      +\phi_c
      -c_0\left[
      \frac{e^{f_{1c}}}{256\alpha}\left(1+\frac{\phi_c}{3 c_0}\,r\right)
      +\frac{\nu_3 }{24} \,r
      \right] \,r  
     +{\cal O}(r^3) \ ,
\label{expphic}  
\end{equation}
where $f_{0c}$, $f_{1c}$, $\nu_3$, $\phi_c$, and $c_0$ are arbitrary constants, 
in terms 
of which all higher-order coefficients can be expressed.
For the dilatonic coupling function $F=\alpha e^{-\gamma\phi}$,
with $\gamma>0$, we assume instead a non-analytic behaviour of the form 
$f_i = f_{ic} + p_i(r)\,e^{-\gamma \frac{c_0}{r}} +\cdots$ for the metric functions,
where $c_0>0$, $p_i(r)$ are polynomials in $r$, and $\cdots$ indicate terms of
order less then $e^{-\gamma \frac{c_0}{r}}$. A similar expression is employed for
the scalar field but with the addition of the singular term $-c_0/r$ as in Eq. (\ref{expphic}).
Then, to lowest power in $e^{-\gamma \frac{c_0}{r}}$, we obtain
\begin{equation}
f_0 =  f_{0c}+
\left(\frac{e^{f_{1c}}}{16\gamma^2 \alpha e^{-\gamma \phi_c}}\right)
 r^2 \left(1-2\frac{r}{\gamma c_0}\right) e^{-\gamma \frac{c_0}{r}}
 \ , \ \ \ 
f_1 = f_{1c} +
\left(\frac{e^{f_{1c}}}{16\gamma^2 \alpha e^{-\gamma \phi_c}}+\nu_3 r\right) r^2\,
e^{-\gamma \frac{c_0}{r}} \ , 
\label{expf1org}
\end{equation}

\vskip -0.5cm
\begin{equation}
\phi =  -\frac{c_0}{r}+\phi_c + c_0 \left[\frac{e^{f_{1c}}}{32 \gamma^2 \alpha e^{-\gamma \phi_c}} \left(1-\frac{3}{2}\,\frac{1}{\gamma c_0}\,r\right) +\frac{\nu_3}{2}\,r\right]
r e^{-\gamma \frac{c_0}{r}}.
\end{equation}

The effective stress-energy tensor (\ref{tmunu}) remains regular over the
entire spacetime and for either form, polynomial or exponential, of the coupling function. 
At large distances, all components vanish in accordance to the asymptotically-flat
limit. At the center ($r=0$), for the quadratic coupling function,
we find the following values
\begin{equation}
T^t_t(0) =3/32\alpha \ \equiv -\epsilon_0 \ , \ \ \ 
T^r_r(0) = T^\theta_\theta(0) = T^\varphi_\varphi(0) = 2/32\alpha
\equiv\  p_0 \,.
\label{Torg}
\end{equation}
Therefore, the energy density $\epsilon_0$ and isotropic
pressure $p_0$, at $r=0$, satisfy
the homogeneous equation of state $p_0= -\frac{2}{3}\epsilon_0$.
Interestingly, neither quantity
depends on the mass or the scalar charge. 
Moreover, the energy density at the center is negative for
positive $\alpha$.
We note that for polynomial coupling functions with powers $n>2$ 
and for the dilatonic coupling function all components of
the stress-energy tensor vanish at $r=0$.

As expected, the spacetime itself remains everywhere regular.
Substitution of the expansion Eqs.~(\ref{expf01c})--(\ref{expphic}) 
in the curvature invariants
yields for the polynomial coupling function with $n=2$
\begin{equation}
R(0) = -9/32\alpha \ , \ \ \ 
R_{\mu\nu} R^{\mu\nu}(0) = 21/(32\alpha)^2 \ , \ \ \ 
R_{\mu\nu\kappa\lambda} R^{\mu\nu\kappa\lambda}(0) = 15/(32\alpha)^2 \ , \ \ \ 
R_{\rm GB}^2(0) =  12/(32\alpha)^2 \ ,
\label{curvinvorg}
\end{equation}
while for higher powers 
$n>2$ and for dilatonic coupling function
the curvature invariants vanish at the center. 

\section{Numerical approach and results}

%
In order to solve the ODEs numerically, we introduce the new coordinate
$\xi = \frac{r_0}{r} $, where $r_0$ is a scaling parameter. 
We note that
the field equations are in fact invariant under the scaling transformation
$r \to \lambda r$ and $F \to \lambda^2 F$, where  $\lambda > 0$.
Hence, we may choose $r_0=1$ without loss of generality.
We treat the ODEs as an initial value problem using the fourth-order
Runge Kutta method.
The initial conditions follow from the asymptotic expansion of Eqs.~(\ref{asym_exp})
\begin{equation}
\left. f_0\right._{\rm ini} =  0 \ , \ \ \ 
\left. f_1\right._{\rm ini} =  0 \ , \ \ \ 
\left. f'_1\right._{\rm ini} =  2M\ , \ \ \ 
\left. \phi\right._{\rm ini} =  \phi_\infty\ , \ \ \ 
\left. \phi'\right._{\rm ini} = -D\,, 
\label{inival}
\end{equation}
where a prime denotes derivative with respect to $\xi$.
We note that all higher-order terms in $\xi$ can be expressed in 
terms of $M$,  $D$ and $\phi_\infty$. 
Thus, for a given coupling function the solutions are completely determined
by these three parameters. For the numerical procedure, we consider 
$\xi_{\rm min} \leq \xi \leq \xi_{\rm max}$,
with $\xi_{\rm min}=10^{-8}$ and $\xi_{\rm max}=100$ 
for the quadratic coupling function,
and  $\xi_{\rm max}=10$ for the dilatonic coupling function.
\begin{figure}[t]
\begin{center}
(a)\mbox{\includegraphics[width=.47\textwidth, angle =0]{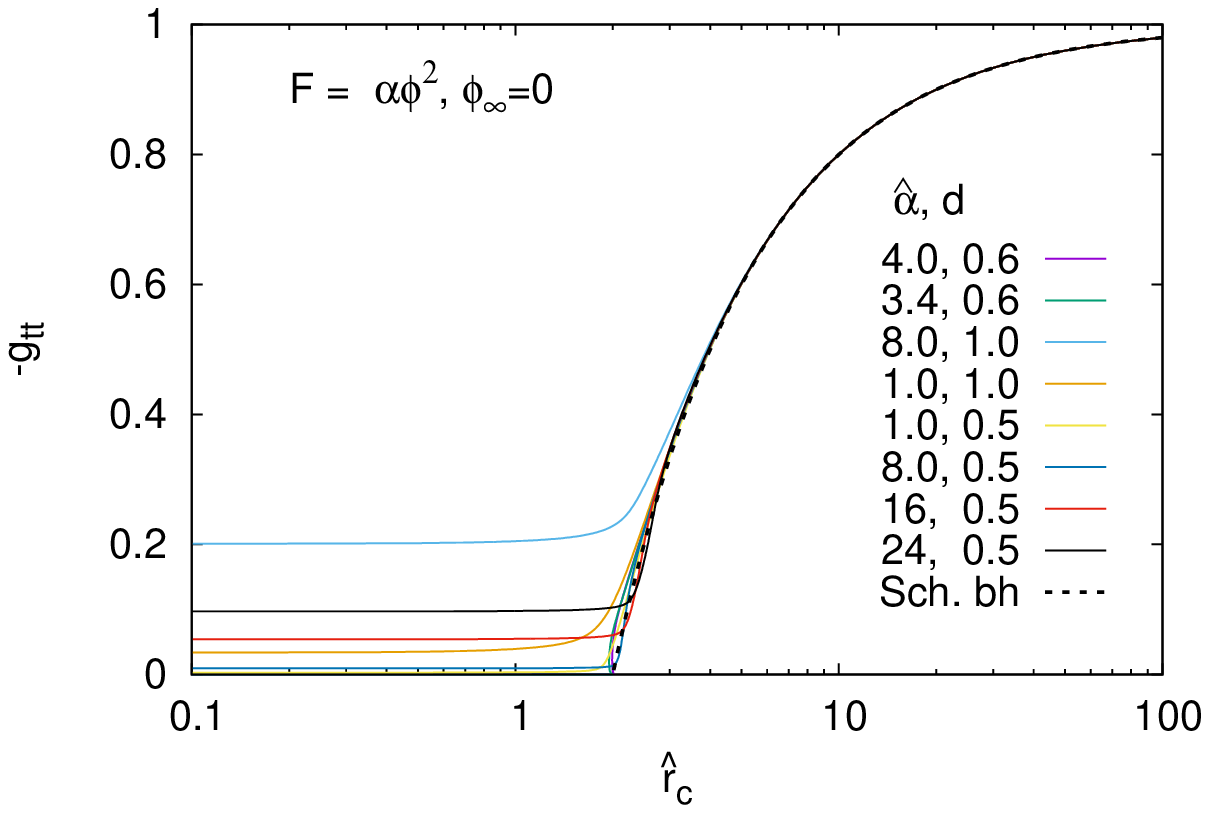}}
(b)\mbox{\includegraphics[width=.47\textwidth, angle =0]{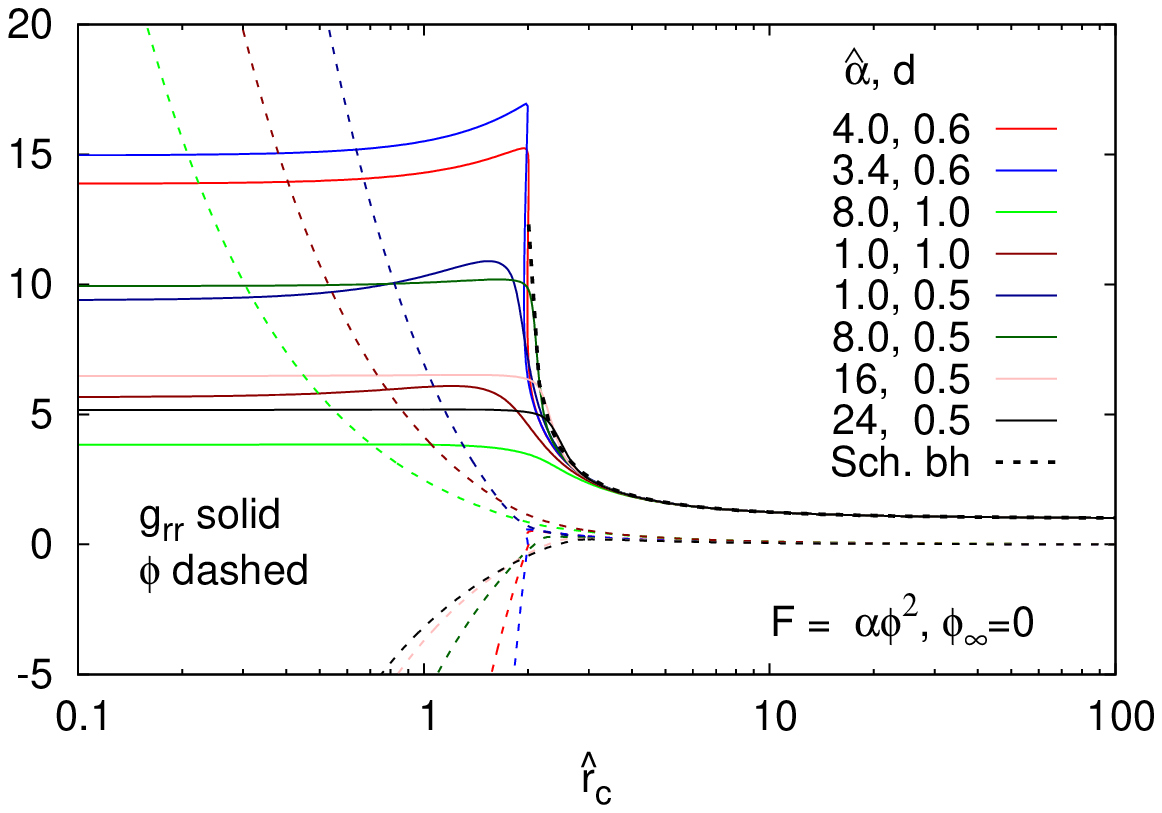}}
\end{center}
\caption{Particle-like solutions for 
the quadratic coupling function with $\phi_\infty=0$:
(a) The metric component $-g_{tt}$
vs the scaled circumferential coordinate $\hat{r}_{\rm c}$;
(b) the metric component $g_{rr}$
and the scalar field $\phi$ 
vs $\hat{r}_{\rm c}$.
\label{solutions}
}
\end{figure}
Typical examples of these particle-like solutions are presented
in Fig.~\ref{solutions}, where we have introduced
the scale invariant quantities $d=D/M$, $\hat{\alpha}=8\alpha/M^2$
and  $\hat{r}_c=r_{\rm c}/M$, with $r_{\rm c}$ the circumferential radius.
They clearly demonstrate the regularity of the metric 
at the center, in full agreement with the asymptotic expansions.

\begin{figure}[t!]
\begin{center}
(a)\mbox{\includegraphics[width=.47\textwidth, angle =0]{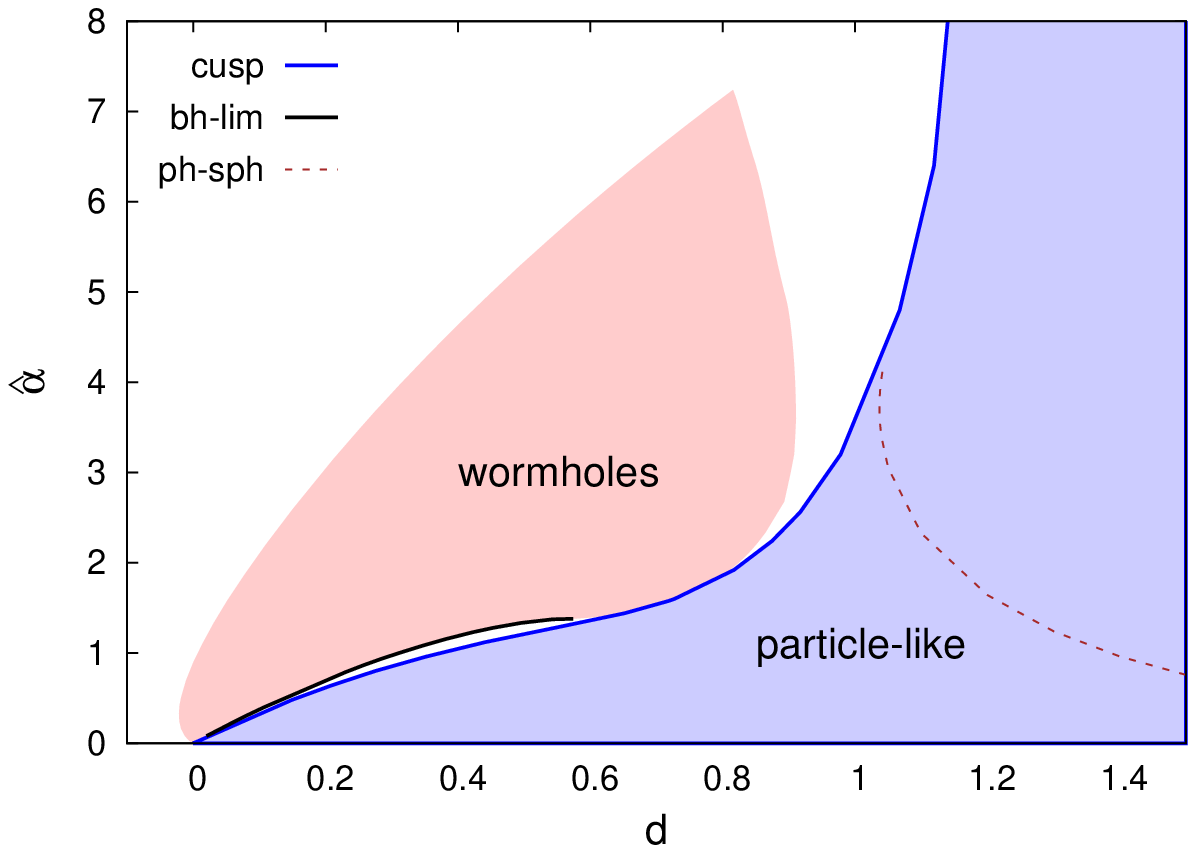}
(b)       \includegraphics[width=.47\textwidth, angle =0]{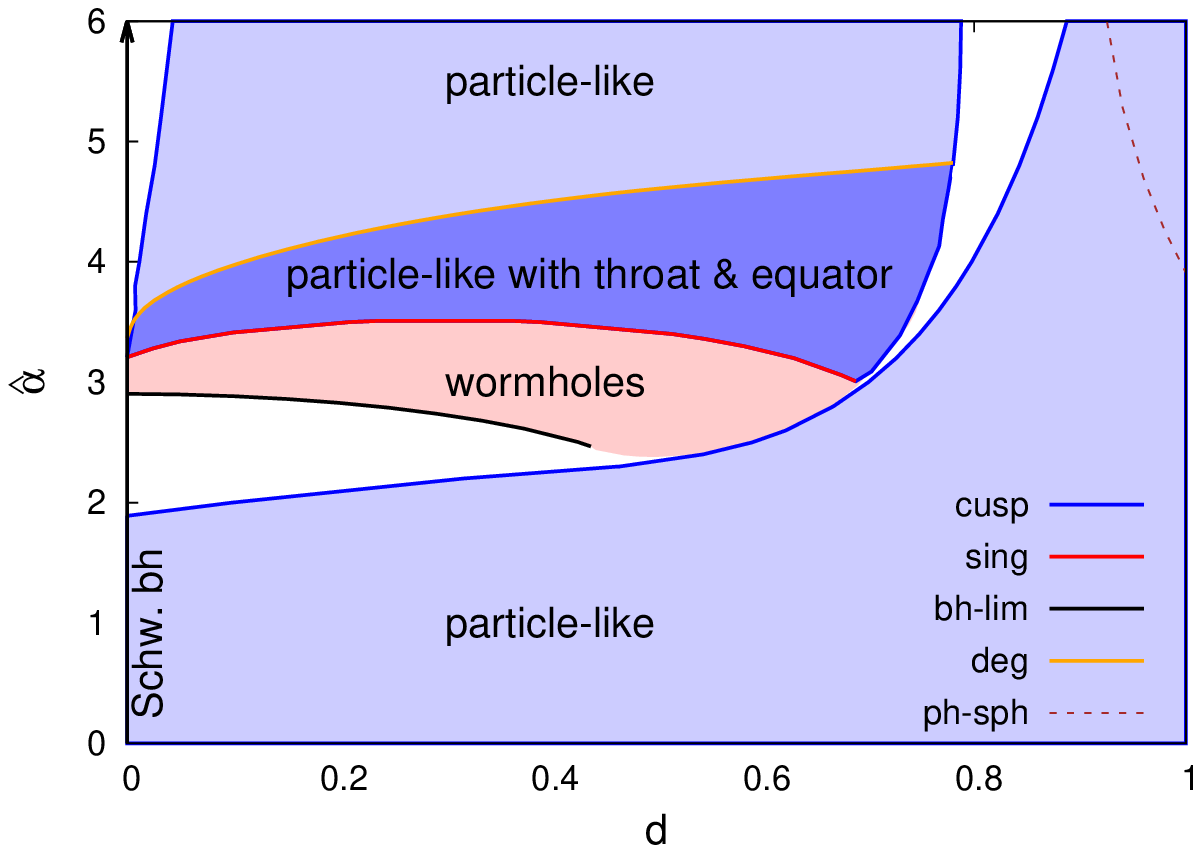}}
\end{center}
\caption{Domains of existence:
The scale invariant quantities $d$ and $\hat{\alpha}$
of particle-like solutions (blue) 
for the (a) dilatonic coupling function
and (b) quadratic coupling function,
with $\phi_\infty=0$.
Also indicated are the wormholes (red) and black hole (black) solutions.
In the dark blue region particle-like solutions and wormholes co-exist.
The dashed curve indicates the boundary of solutions for which
lightrings exist. Solutions without lightrings are located above and 
to the right of the dashed curve.
\label{domains}
}
\end{figure}

%
%
In delimiting the domain of existence of these particle-like 
solutions, we again employ the 
scale invariant quantities $d$ and $\hat{\alpha}$,
restricting to $\alpha\geq 0$. The respective domains
are exhibited in Fig.~\ref{domains}a for the dilatonic coupling function
and in Fig.~\ref{domains}b for the quadratic coupling function, with $\phi_\infty=0$.
For the dilatonic coupling function, particle-like solutions exist 
when, for some fixed value of $\hat{\alpha}$,
the scaled scalar charge $d$ is larger 
than a critical value $d_{\rm cr}(\hat{\alpha})$. 
The curve $d_{\rm cr}(\hat{\alpha})$ then
forms the boundary of the domain of existence. 
In the limit $d \to d_{\rm cr}$
the solutions develop a cusp singularity.
We note that for the numerical procedure we have to diagonalize
the Einstein and scalar-field equations with respect to the
second derivatives. This introduces the determinant of the
coefficient matrix in the denominator of the diagonalized ODEs.
A cusp singularity emerges if the determinant develops a zero
at some point $r_c$. As a consequence the second derivatives
either diverge or possess a jump at $r_c$. In our numerical procedure
we monitor the determinant and stop the computation if it changes sign.

For the quadratic coupling function, the domain of existence is bounded
by solutions with cusp singularities
and by singular solutions where $g_{tt}$ vanishes at some point.
For a non-vanishing asymptotic value of the scalar field,
i.e. $\phi_\infty=1$, the domain of existence consists of two disconnected
regions. For $\phi_\infty=0$, the domain of existence has a
more interesting structure. Due to
the symmetry $\phi \to -\phi$, all particle-like solutions come now in
pairs with positive and negative scalar charge -- 
Fig.~\ref{domains}b is restricted to positive values of $D$.
Also, the domain of existence consists of many disconnected regions
due to the emergence of excited solutions for large $\hat{\alpha}$
(only the first two regions are shown in Fig.~\ref{domains}b).
In some region of the domain of existence, independently of
the value of $\phi_\infty$, particle-like solutions appear where
the spacetime around them possesses a throat and an equator.

Let us compare with the domain of existence of wormholes and black holes.
For the dilatonic coupling function, the domains of existence of particle-like
solutions and wormholes do not overlap, although their boundaries almost touch.
This is in contrast to the case of the quadratic coupling function. Here, the region
where particle-like solutions with a throat and an equator exist overlaps with the
domain of existence of wormholes. 
Indeed, in this region the wormholes can be constructed
from the particle-like solutions by cutting the spacetime at the throat
or equator and continuing symmetrically to the second asymptotic region
after a suitable coordinate transformation. 
The black-hole solutions form part of the boundary of the domain of existence
of the wormholes, but not of the particle-like solutions.

\section{Relation to Fisher solution}

%
For vanishing coupling function, EsGB theory reduces to 
GR with a self-gravitating scalar field.
Here (singular) solutions are known in closed form 
\cite{Fisher:1948yn}, 
\cite{Janis:1968zz,Wyman:1981bd,Agnese:1985xj,Roberts:1989sk}.
In isotropic coordinates, as in Eq.~(\ref{met}), these GR solutions read 
\begin{equation}
e^{f_0} = \left(\frac{1-r_{s}/r}{1+r_{s}/r}\right)^{2s} \ , \ \ \
e^{f_1} = \left(1-r_{s}^2/r^2\right)^2\, e^{-f_0} 
            \ , \ \ \
\phi=\phi_\infty \pm \frac{d}{2} \, f_0 ,
\label{Fishersol}
\end{equation}
where $d=D/M$ is again the scaled scalar charge and $s=1/\sqrt{1+d^2/4}$.
The singularity is located at $r_{s} =M/2s$.
The Schwarzschild black hole is obtained in the limit $d\to 0$, 
corresponding to $s\to 1$.

The EsGB solutions can be related to the aforementioned GR solutions as follows.
In the limit $d \to \infty$, the scale symmetry of the field equations
implies the scaling of mass, scalar charge,
and coupling strength as
$M \to \lambda M$, $D \to \lambda D$ and  $\alpha \to \lambda^2 \alpha$.
When scaling with $\lambda = 1/D$,
the coupling functions vanish in this limit,
and we are left with a self-gravitating scalar field in GR.
This solution can be obtained from the Fisher solution Eq.~(\ref{Fishersol})
in the limit $d\to \infty$ for fixed $r_{s}$, which implies
$s\to 0$ with fixed $s\, d$.
Similarly, in the limit $\hat{\alpha} \to 0$, for fixed $d$ the
particle-like solutions tend to the Fisher solution,
if $r$ is larger than the coordinate of the Fisher singularity, 
i.e.  $r> r_{\rm s}$.
%
%

\section{Observational effects}

%
A photon emitted by a source located at the center is seen by an
observer in the asymptotic region with a redshift factor
$z=\lambda_{\rm asym}/\lambda_{\rm emit}-1$, where 
$\lambda_{\rm asym, emit}$ is its wavelength at detection and emission,
respectively. This quantity depends on the metric component $g_{tt}(0)$ at
$r=0$, 
and is given by the expression $ z= e^{-f_0(0)/2}-1$\,.
For the dilatonic and quadratic coupling functions with
$\phi_\infty=1$, the redshift factor increases with decreasing $\hat{\alpha}$,
and takes very large values as the Fisher limit is approached.
Interestingly, for particle-like solutions arising for the
quadratic coupling function with $\phi_\infty=0$, 
the redshift factor can take both positive and negative
values, and thus allow also for a gravitational blue shift.

\begin{figure}[t!]
\begin{center}
(a)\mbox{\includegraphics[width=.47\textwidth, angle =0]{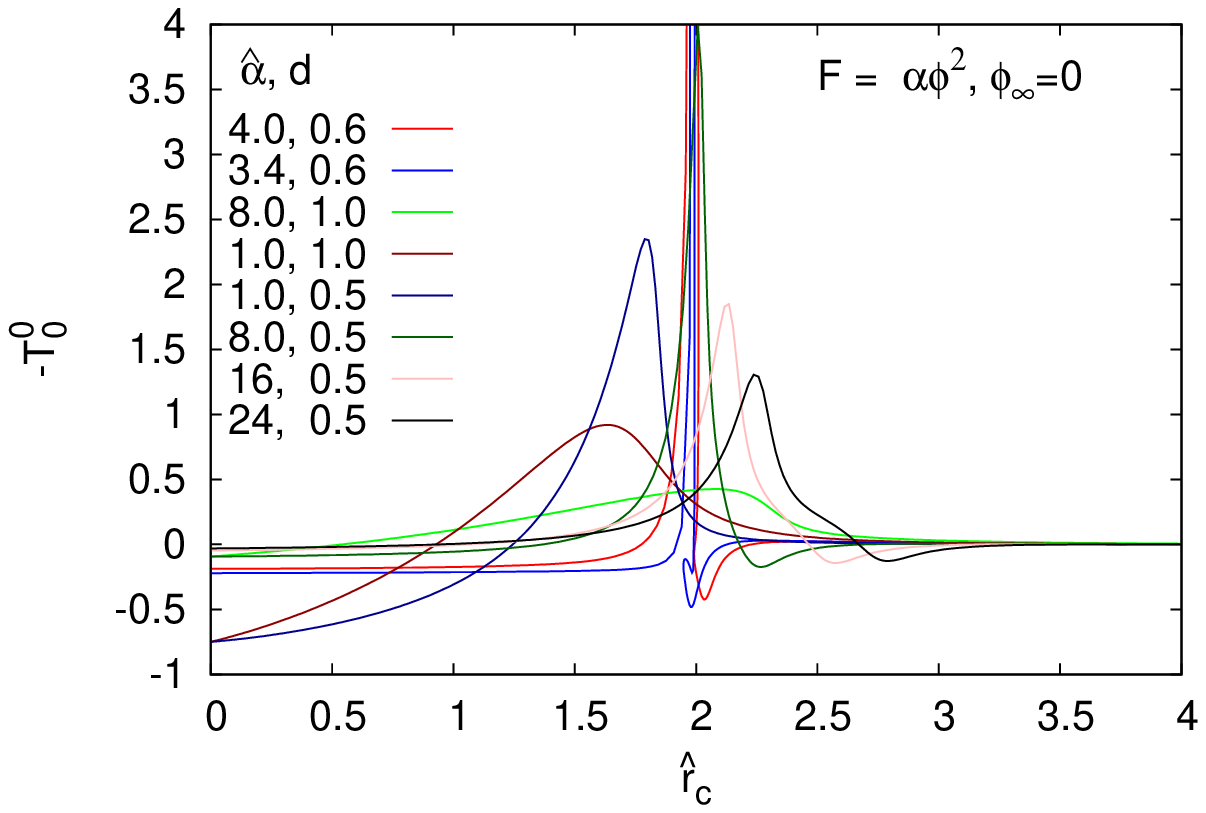}
(b)      \includegraphics[width=.47\textwidth, angle =0]{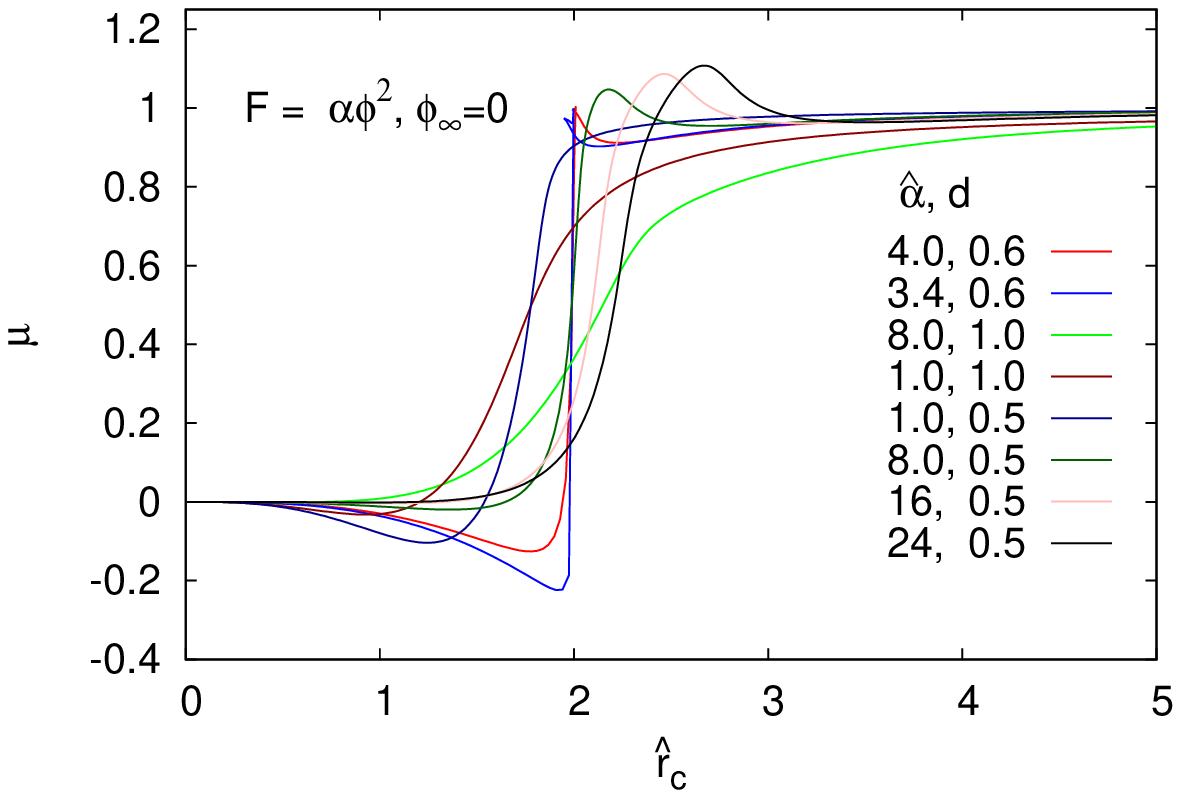}}
\end{center}
\caption{
(a) The energy density $\rho=-T_0^0$ vs the circumferential radius
$\hat r_c$, and
(b) the mass function $\mu(\hat r_{\rm c})$ within a volume with 
circumferential radius $\hat r_{\rm c}$, 
for the quadratic coupling function ($\phi_\infty=0$).
\label{mass}
}
\end{figure}

Let us now consider the energy density $\rho=-T^0_0$ and
the mass function $\mu(\hat r_{\rm c})$
defined via 
$\mu = - 1/2 \int_0^{\hat r_{\rm c}} T^0_0 \hat r_{\rm c}^2 d \hat r_{\rm c}$
of these particle-like solutions, 
where $\hat r_{\rm c}$ is the circumferential radius.
These are illustrated in Fig.~\ref{mass}
for the quadratic coupling function ($\phi_\infty=0$).
As expected from the shell-like behaviour of the
energy density, shown in Fig.~\ref{mass}a,
the mass function $\mu(\hat r_{\rm c})$,
shown in Fig.~\ref{mass}b, exhibits a
characteristic steep rise towards
its asymptotic value in the vicinity of $\hat r_{\rm c}=2$
for many of the particle-like solutions,
qualifying them as
highly compact objects. 
This holds for both the quadratic and the dilatonic coupling function.

\begin{figure}[t!]
\begin{center}
(a)\mbox{\includegraphics[width=.47\textwidth, angle =0]{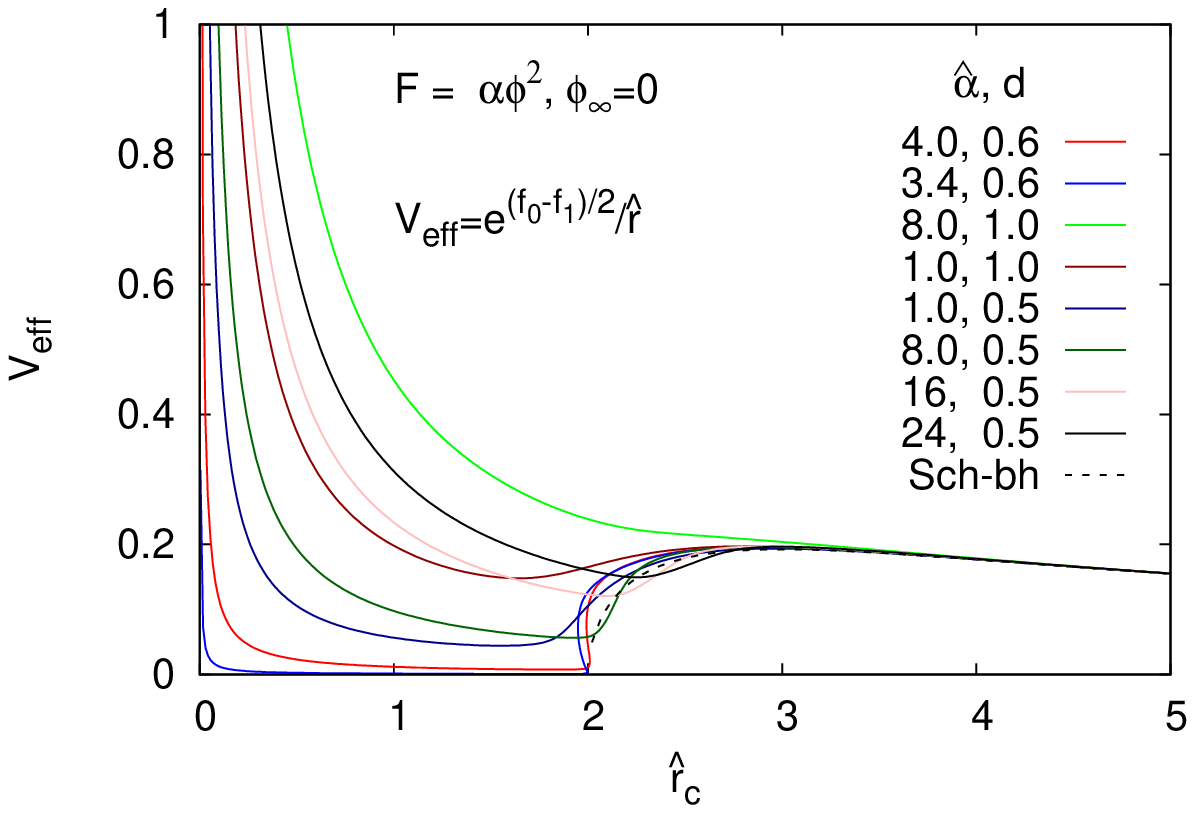}}
(b)\mbox{\includegraphics[width=.47\textwidth, angle =0]{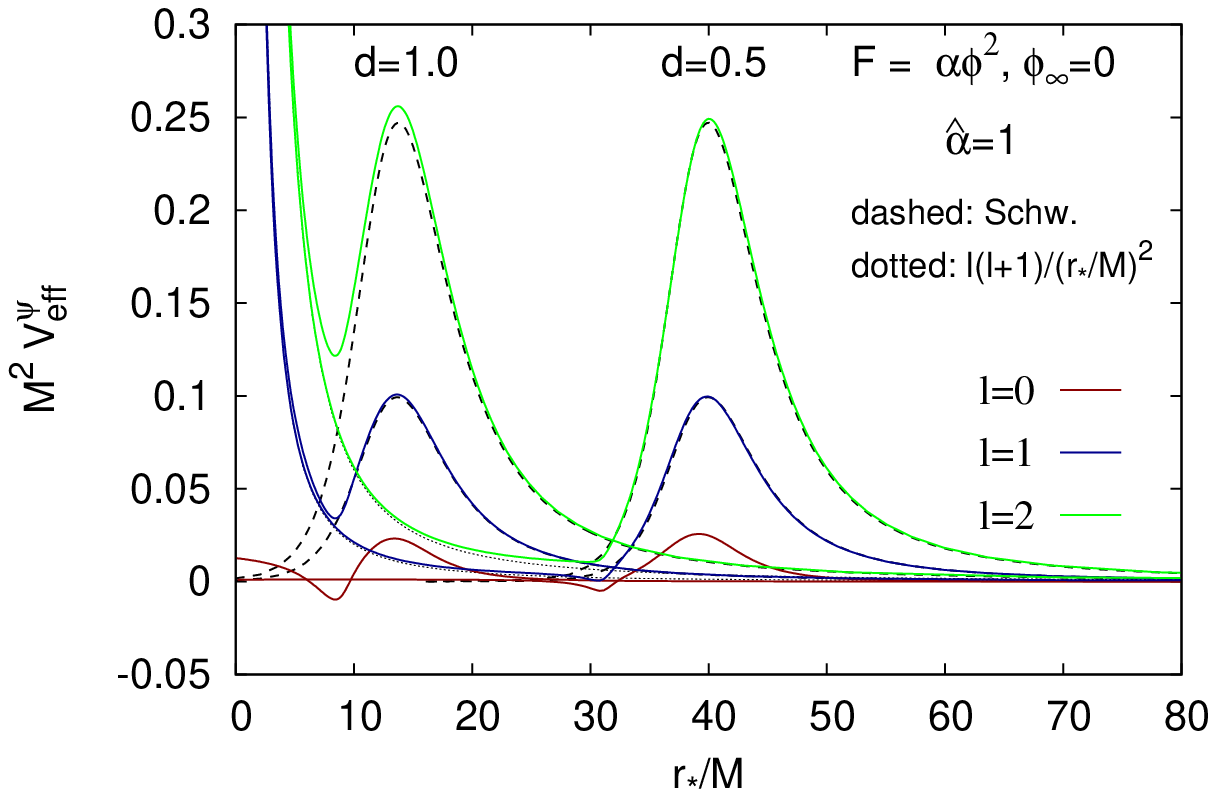}}
\end{center}
\caption{(a) Effective potential $V_{\rm eff}$ 
and lightrings located at the extrema of $V_{\rm eff}$
vs the circumferential radius $\hat r_{\rm c}$ of particle-like solutions
for the 
quadratic coupling function with $\phi_\infty=0$.
(b) Effective potential $V^\psi_{\rm eff}$ for a test scalar particle with
$l=0,1,2$ moving in the background of two indicative particle-like solutions. 
The dashed curves stand for the corresponding Schwarzschild limit.
\label{veff}
}
\end{figure}

An inspection of the lightlike geodesics in the gravitational background of our
particle-like solutions reveals that these may
also possess lightrings, 
and thus represent UCOs \cite{Cardoso:2017cqb}.
To start with, the independence of the Lagrangian
${\cal L}=g_{\mu\nu} \dot x^\mu \dot x^\nu/2$ of the $(t,\varphi$) coordinates
leads to two conserved quantities, namely the energy $E=e^{f_0} \dot t$ and
the angular-momentum $L=e^{f_1} r^2 \dot \varphi$ of the particle. Employing
these two expressions in the geodesic equation $(ds/d\tau)^2=0$ for photons
moving in the equatorial plane, we find
\begin{equation}
e^{f_0+f_1} \dot r^2=(E+L V_{\rm eff})(E-L V_{\rm eff}),
\label{Veff_photons}
\end{equation}
where $V_{\rm eff}=e^{(f_0-f_1)/2}/r$.
In Fig.~\ref{veff}a, the aforementioned effective potential for lightlike geodesics 
is shown for the same set of solutions as in Fig.~\ref{mass}.
Depending on the parameters, the effective potential
possesses either no lightring or a pair of lightrings,
associated with the extrema of $V_{\rm eff}$.
Photons with energies smaller than the value of $L\,V_{\rm eff}$ at its
local maximum remain in bound orbits around the particle-like solutions. 
The areas in the domains of existence where solutions with 
lightrings emerge are indicated in Fig.~\ref{domains} by the dashed curves.

The presence of a pair of lightrings is
in accordance with a theorem on lightrings of UCOs,
showing that there should indeed be two lightrings
for a  smooth spherically symmetric metric,
with one of the lightrings being stable
\cite{Cunha:2017qtt}.
As pointed out in \cite{Keir:2014oka,Cardoso:2014sna}
the presence of a stable lightring
may possibly lead to nonlinear spacetime instabilities.
Clearly, if such instabilities would indeed be present,
astrophysical observations of such UCOs would require 
their lifetime to be large on relevant astrophysical timescales.

When considering the propagation of gravitational waves around these
UCOs, the absence of a horizon will lead to a sequence of echoes
with decreasing amplitude \cite{Cardoso:2017cqb,Cardoso:2016rao}.
We may use the equation of motion of a free, test scalar
particle $\Psi$ in order to demonstrate this. Assuming that 
$\Psi(t,r,\theta,\varphi)=\psi(t,r)\,e^{-f_1/2} Y^m_l(\theta, \varphi)/r$,
its equation $\partial_\mu (\sqrt{-g}\,\partial^\mu \Psi)=0$ takes the
wave form $(\partial_t^2- \partial^2_{r_*} +V^{\psi}_{\rm eff})\,\psi=0$,
where the effective potential reads
\begin{equation}
V_{\rm eff}^{\psi}=e^{f_0-f_1} \left[\frac{l(l+1)}{r^2}+ \frac{2 (f_1'+f_0') +
r f_1'f_0'+2r f_1 ''}{4r}\right],
\end{equation}
in terms of the tortoise coordinate $r_*= \int_0^r e^{(f_1-f_0)/2} dr$. 
The profile of $V_{\rm eff}^{\psi}$ is shown in Fig. \ref{veff}b for two of our particle-like
solutions and for $l=0,1,2$ for the test scalar field. We note that, for $l>0$,
$V_{\rm eff}^{\psi}$ diverges at $r=0$, thus creating an angular-momentum
barrier. All modes of an incoming test scalar field will be partially transmitted through
the finite local barrier and partially reflected back to infinity. However, the transmitted
modes with $l >0$ will undergo a perpetual process of full and partial reflection 
between the angular-momentum and local barrier, respectively. This process
will lead to an infinite number of echoes with a decreasing amplitude every time the
wave gets partially transmitted through the local barrier as it moves outwards.

\noindent{\textbf{~~~Acknowledgement.--~}}
BK and JK gratefully acknowledge support by the
DFG Research Training Group 1620 {\sl Models of Gravity}
and the COST Action CA16104. BK and PK acknowledge helpful
discussions with Eugen Radu and Athanasios Bakopoulos, respectively.

\end{document}